\documentclass[conference]{IEEEtran}

\usepackage{xspace}
\usepackage{listings}
\usepackage{xcolor}
\usepackage{balance}
\usepackage{subcaption}
\usepackage[inline]{enumitem}
\usepackage{hyperref}
\usepackage{graphicx}
\usepackage{booktabs}

\newcommand\ddef{& ::= &}

\newcommand\lsep{\\&|&}
\newcommand\esep{\\&&}

\usepackage{ifthen}
\newboolean{showcomments}
\setboolean{showcomments}{true} 
\ifthenelse{\boolean{showcomments}}
{\newcommand{\nbnote}[3]{
  \fcolorbox{gray}{yellow}{\bfseries\sffamily\scriptsize#1} 
  {\color{#2} \sffamily\small$\blacktriangleright$\textit{#3}$\blacktriangleleft$}
  }
}
{\newcommand{\nbnote}[3]{}
 
}

\newcommand\ostrich{\texttt{OSTRICH}\xspace}
\newcommand\chameleon{\texttt{CHAMELEON}\xspace}

\newcommand\kw[1]{\texttt{\small #1}}
\newcommand\widget[1]{\textit{#1}}

\newcommand\cf{cf.\xspace}

\def\BibTeX{{\rm B\kern-.05em{\sc i\kern-.025em b}\kern-.08em
    T\kern-.1667em\lower.7ex\hbox{E}\kern-.125emX}}
\begin{document}

\title{{CHAMELEON}: OutSystems Live Bidirectional Transformations
}

\author{\IEEEauthorblockN{Hugo Lourenço}
\IEEEauthorblockA{\textit{OutSystems} \\
hugo.lourenco@outsystems.com
}
\and
\IEEEauthorblockN{João Costa Seco}
\IEEEauthorblockA{\textit{NOVA University Lisbon} \\
joao.seco@fct.unl.pt
}
\and
\IEEEauthorblockN{Carla Ferreira}
\IEEEauthorblockA{\textit{NOVA University Lisbon} \\
carla.ferreira@fct.unl.pt
}
\and
\IEEEauthorblockN{Tiago Simões}
\IEEEauthorblockA{\textit{OutSystems} \\
tiago.simoes@outsystems.com}
\and
\IEEEauthorblockN{Vasco Silva}
\IEEEauthorblockA{\textit{OutSystems} \\
vasco.silva@outsystems.com}
\and 
\IEEEauthorblockN{Filipe Assunção}
\IEEEauthorblockA{\textit{OutSystems} \\
filipe.assuncao@outsystems.com}
\and
\IEEEauthorblockN{André Menezes}
\IEEEauthorblockA{\textit{OutSystems} \\
andre.menezes@outsystems.com}
}


%

%
%
%
%


\maketitle

\begin{abstract}
In model-driven engineering, the bidirectional transformation of models plays a
crucial role in facilitating the use of editors that operate at different levels
of abstraction. This is particularly important in the context of
industrial-grade low-code platforms like OutSystems, which feature a
comprehensive ecosystem of tools that complement the standard integrated
development environment with domain-specific builders and abstract model
viewers.

We introduce CHAMELEON, a tool that enables the dynamic definition of a live
bidirectional model transformation in a declarative manner by leveraging simple
and intuitive component patterns. Through this approach, we can gradually define
the view and synthesis paths to an abstract model built on top of a low-code
metamodel.

We devise a standard parser-generating technique for tree-like models that
builds upon extended grammar definitions with constraints and name binders.
We allow for a greater overlap of model patterns that can still be disambiguated
for a clear lens-like behaviour of the transformation.

CHAMELEON is evaluated in the fragment of the OutSystems language targeting the
definition of user interfaces. To assess performance we used a large set of real
OutSystems applications, with approximately 200K UI widgets, and a database of
curated widget patterns. We found a worst-case processing time of 92ms for
complete models in our benchmark, which is still suitable for the operation of
an interactive model editor.
\end{abstract}




\begin{IEEEkeywords}
low-code, model transformation, virtualization
\end{IEEEkeywords}

\section{Introduction}


The holy grail of low-code frameworks is to serve a wide set of skilled
professionals. Low-code allows non-expert developers to build software with a
smooth learning curve, and  provide experienced developers with a more
productive environment. 

Low-code platforms usually work in a specific domain by leveraging the full
power of a given platform or domain-specific language (DSL). However, the
adoption of a DSL or low-code approach is many times accompanied by a limited
expressive power~\cite{Vis08}, thus forcing a set of programming patterns onto
the developers. Model-based approaches and DSLs with unidirectional model
transformations typically have a write-once nature where changes can only be done at
the higher abstract level. When changes can also be performed on a more concrete
level, they get lost if the higher abstract level evolves.
%
In a scenario where an abstract representation is shown to the users of the
higher abstract levels, they are usually kept in the dark about the parts of the
concrete model not captured in the more abstract model.
In prior work, we defined a bidirectional approach to ``application
builders''~\cite{ExBuilder,WFBuilder,IBuilder} that does not have this
write-once limitation~\cite{DBLP:conf/models/RamalhoLS21}, but still ignores the parts of the concrete model that
application builders do not produce and do not fit the defined abstractions. 
%
The approach we followed in~\cite{DBLP:conf/models/RamalhoLS21} is to define
fixed transformations between abstract models. The transformations are specific
to each ``application builder'' and the actual OutSystems application metamodel.
For example, there is a builder for the design of business processes~\cite{WFBuilder} and a
builder that uses existing application screens to define the navigation of an
application~\cite{ExBuilder}.  
Such an approach is only suitable for an asynchronous editing process where a
model is loaded onto a builder, modified, and then transformed back into the
original metamodel.


In this paper, we present a multi-layered bidirectional model-based design that
efficiently abstracts the concrete (native) model using a gradual approach. 
%
%
%
Although our approach can be applied to any part of the model, this paper
focuses on the user interface, where concrete components are called (native) widgets and
abstract components are called virtual widgets.
We start with the current OutSystems application model and abstract a set of
widget patterns, progressively defined by experts in a back-office application
of the platform. 
The abstract model presented to the user, in a virtual widget tree, is considerably smaller
than the native  widget tree but, nevertheless, captures all the components that
are not matched to any specified virtual widget pattern. This set of
components left unmatched and not abstracted, are not hidden away, but remain
visible to the user in the form of properly bounded native widgets.
Our approach thus supports a gradual model transformation that allows the definition
of new virtual widget patterns to capture more and more parts of the concrete
model. Step by step, the complex widget tree of the native model is abstracted
into a more intuitive and simpler virtual widget tree.

We specially focus on the efficiency of the parsing process to allow for changes
in the application model to be immediately seen at the abstract level and
vice-versa. We use parsing generator techniques to efficiently scan and abstract
a tree-like model. Unlike traditional bidirectional transformation engines that
work by querying single graph-based patterns on a model one by one~\cite{kahani2019,DBLP:journals/sosym/GreenyerK10},
we use a parsing
approach that tackles all patterns in a single pass.
By improving on an existing, and slower, detection approach, we aim at having a
live editing experience of the model that can support changes at different
levels of abstraction. 

\chameleon improves state-of-the-art virtualization techniques at
OutSystems that take several tens of seconds to load and process a typical application
model. In comparison, the new virtualization process takes less than 100 ms per application model and less than
10 ms per application screen. Furthermore, our approach can operate in smaller
increments, parsing and showing only the screen that is currently relevant to
the user. 
%
The overall objective is to support model editors to operate live at
different levels of abstraction while covering the entirety of the low-code
metamodel, and allow model visualization and edition by diverse developer
profiles.



We introduce a metamodel for the virtual widgets that simultaneously
defines the patterns to be captured in the abstract layer and
the synthesis process for virtual widgets into the native model.
We use component identifiers to bind different definitions and express
the connection between properties of the native and abstract models.
From the virtual widgets definition, we generate an attribute grammar
with conditions~\cite{parr1993use,antlr}.
A virtual widget may consist of several different patterns, for instance a "Boolean Input" virtual widget can be represented either as a checkbox, a switch, or a group of two buttons representing the True and False alternatives.
Pattern definitions may overlap - two grouped buttons may also be seen as a specific case of an "Enum Input" used to select between a fixed set of alternatives.

The grammar produced is inherently ambiguous, which we tackle by
building a custom-defined generalized LR parser~\cite{tomita87}. 
Noticeably, Tomita's GLR parsers~\cite{tomita87} explore all alternatives in parallel,
but we explore  alternatives in sequence giving priority to shift-reduce
conflicts to obtain the longest sequence (ensuring a higher level of virtualization). 
Our implementation incorporates
backtracking as a way of recuperating from parsing errors and following to the
next alternative.

To capture the substitution process of native widgets by virtual widgets, the original (native)
metamodel is dynamically extended with additional classes to represent virtual widgets. 
The abstract metamodel is thus a proper extension of the native metamodel
as it can fully represent native widget applications. 
Editors of the abstract model can still represent and preview native widgets, by
using a complete widget tree and a preview functionality for native widgets,
respectively. 


We evaluate \chameleon using a benchmark with 239 application modules and
approximately 4K screens containing around 200K UI widgets. We used a database
of 106 patterns grouped in 31 virtual widgets. We found linear trends in both the
efficiency and effectiveness of the approach when considering full application
modules. In practice, the parsing of one screen constitutes a negligible
constant pass in the virtualization of individual screens with parsing times
smaller than 10 ms, which enables its integration in an interactive and
collaborative editor of abstract models.
%



In summary, the main contributions of this work are:

\begin{itemize}

  \item The system architecture, described in \autoref{sec:architecture}, that
allows the gradual definition of new patterns with dynamic extensions to the
original metamodel and allows model editions at multiple levels of abstraction.

  \item A language to define virtual widgets that simultaneously defines patterns
to be captured and the synthesis process for new components, also described in
\autoref{sec:example}.

  \item A multi-layered bidirectional model-based design that efficiently
abstracts the native model in a gradual approach. This means that the parsing
process is a best-effort process that uses all the virtual widgets definitions
available (\autoref{sec:parsing})

  \item A synthesis technique that allows the
interchangeable use of the widget patterns of a virtual widget (\autoref{sec:synthesis}).

\end{itemize}

We finalize the paper by using a benchmark in \autoref{sec:evaluation} to
evaluate our approach and presenting a list of open problems that we leave for
future work (\autoref{sec:futurework}).

In conclusion, the measure of impact is given by the fact that it is in the company's
roadmap to be integrated into its main product. 


\section{System Architecture}\label{sec:architecture}

\begin{figure*}
    \centering
    \includegraphics[width=0.7 \linewidth]{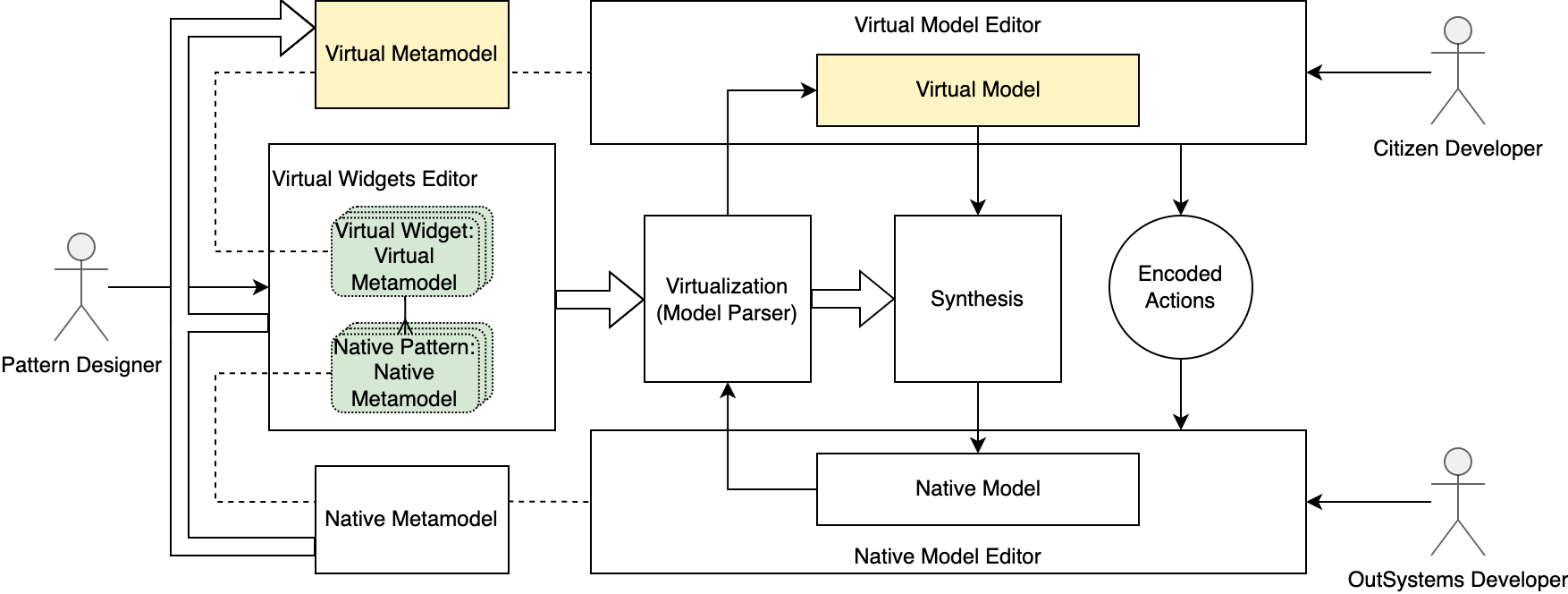}
    \caption{Overview of the proposed approach}
    \label{fig:overview}
\end{figure*}

The components of the system are shown in the overall architecture diagram
(\autoref{fig:overview}) and cover the functionalities for three different
personas: an \textit{OutSystems Developer} that uses the current development
environment and the native model to create applications; a \textit{Pattern
Designer} that uses the native model to define patterns to group and parametrize
them to define virtual widgets; and the \textit{Citizen
Developer}~\cite{CitizenDev,CitizenDev2}. Citizen Developers, central to our
work, create applications using virtual widgets. 
%
They use a virtual model editor that presents a simplified but loss-free widget
tree that allows the editing and creation of new virtual widgets instances, the
read-only preview of native widgets (part of the virtual metamodel), and
also makes available selected operations that transform the native model
directly and are encoded into the editor. Such low-level hard-coded operations
are available for pragmatic and historical reasons to support basic construction
operations in the virtual model editor that do not concern the creation of
widgets (e.g. changing the data model in predefined ways.)

The virtual metamodel is generated having as input the definition of the native
metamodel and the set of virtual widgets defined by the pattern designer. This
is represented by the thick arrow on the left-hand side of the diagram. The
virtual metamodel is then used by the virtual model editor to support its
interactions. As expected, the native metamodel is the basis for the native
model editor operations. The virtual widgets definitions are also used to
generate the virtualization (parsing) and synthesis operations.

In \autoref{sec:example} we present a case study that will support the
detailed description of our parsing method (described in \autoref{sec:parsing}) and
the synthesis process (in \autoref{sec:synthesis}).

\section{Case Study: UI Patterns}\label{sec:example}

\begin{figure}
\centering
    \includegraphics[width=.55 \linewidth]{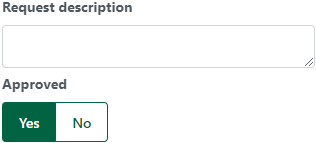}
    \caption{Request edit form}
    \label{fig:form}
\end{figure}

\begin{figure*}
    \centering
    \includegraphics[width=.65 \linewidth]{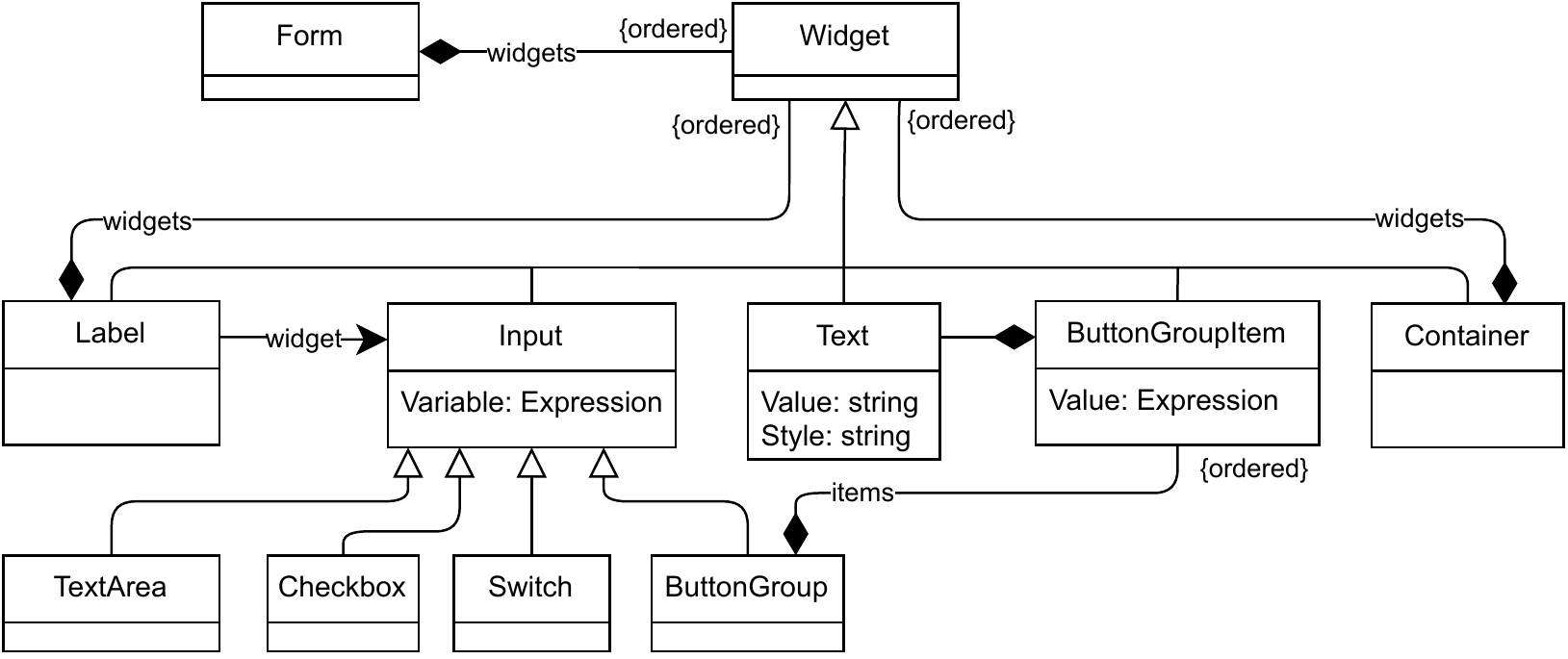}
    \caption{Widgets Metamodel}
    \label{fig:native_metamodel}
\end{figure*}

Our work focuses on simplifying the construction of user interfaces via the
introduction of higher-level constructs that are simpler to manipulate and
understand than the underlying model, the low-code OutSystems application model.
In this section, we present a running example containing a simple form to edit a request in an imaginary corporate application (\autoref{fig:form}). The form allows the introduction of the description of the request along with approval information. In \autoref{fig:native_metamodel} we present the metamodel for defining forms, which is a subset of the actual OutSystems metamodel.
A \widget{Form} consists of a tree of \widget{Widgets}.
\widget{Input} widgets are used to obtain data from the user and bind it to a state variable in the low-code application. \widget{Input} widgets are specialized according to the data type: type \widget{TextArea} is used for textual data, types \widget{Checkbox} and \widget{Switch} are tailored to boolean data, and type \widget{ButtonGroup} is used for multiple alternatives, each represented by a \widget{ButtonGroupItem} widget.
The widget \widget{Container} is a layout widget used to group other widgets. Widget \widget{Label} is used for accessibility purposes, describing an associated input widget.

\begin{figure}
    \centering
    \begin{subfigure}[b]{0.2\textwidth}
        \centering
        \includegraphics[width=.9 \linewidth]{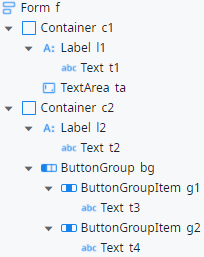}
        \caption{Native tree}
        \label{fig:native_tree}
    \end{subfigure}
    \hfill
    \begin{subfigure}[b]{0.2\textwidth}
        \centering
        \includegraphics[width=.75 \linewidth]{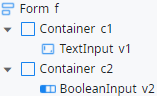}
        \caption{Virtual tree.}
        \label{fig:virtual_tree}
    \end{subfigure}
    \caption{Widget trees}
    \label{fig:trees} 
\end{figure}

\begin{figure}
    \centering
    \includegraphics[width=.55 \linewidth]{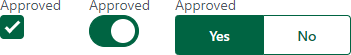}
    \caption{Three ways of representing a boolean input}
    \label{fig:boolean_inputs}
\end{figure}


In \autoref{lst:native_model} we present the model of \autoref{fig:form}, and in \autoref{fig:native_tree} we depict the visual representation of the widget tree. For reasons which will become clear in the next section, we opted to present all models in XML notation.


In our example, the input "Approved" in the form is used to obtain boolean data.
The developer chose a user-friendly representation \widget{ButtonGroup} widget
with two alternatives (\widget{Yes} or \widget{No}), but it would also be possible to use either a
\widget{Checkbox} widget or a \widget{Switch} widget. This simple example
illustrates that, in general, there is more than one equivalent way to represent
the same information (see \autoref{fig:boolean_inputs}).

\ \ \begin{minipage}{0.95\linewidth}
\begin{lstlisting}[basicstyle=\footnotesize\ttfamily,language=XML, caption={Form model}, label={lst:native_model}]
<Form Id="f">
	<Container Id="c1">
		<Label Id="l1" Widget="ta">
			<Text Id="t1" Style="bold" Value="Request description"/>
		</Label>
		<TextArea Id="ta" Variable="Request.Description"/>
	</Container>
	<Container Id="c2">
		<Label Id="l2" Widget="bg">
			<Text Id="t2" Style="bold" Value="Approved by Manager"/>
		</Label>
		<ButtonGroup Id="bg" Variable="Request.Approved">
			<ButtonGroupItem Id="g1" Value="true">
				<Text Id="t3" Value="Yes"/>
			</ButtonGroupItem>
			<ButtonGroupItem Id="g2" Value="false">
				<Text Id="t4" Value="No"/>
			</ButtonGroupItem>
		</ButtonGroup>
	</Container>
</Form>
\end{lstlisting}
\end{minipage}

A quick inspection of \autoref{lst:native_model}, containing the native model for the example form, reveals a lot of repetition and unnecessary detail that the developer should not be concerned about. Each of the two inputs follows the same basic pattern: a \widget{Label} containing a \widget{Text} widget with a descriptive text, followed by an \widget{Input} widget of arbitrary complexity bound to a variable.
We can reduce the developer's effort by introducing higher-level elements, which we denote \textit{virtual widgets}, with a simplified \textit{API}: the developer should only have to decide which variable to bind to. The specific widget representation can then be chosen from a set of alternatives, manually or automatically, and, when needed, the initial value for the label text can be inferred from the variable.

\begin{figure*}
    \centering
    \includegraphics[width=.95 \linewidth]{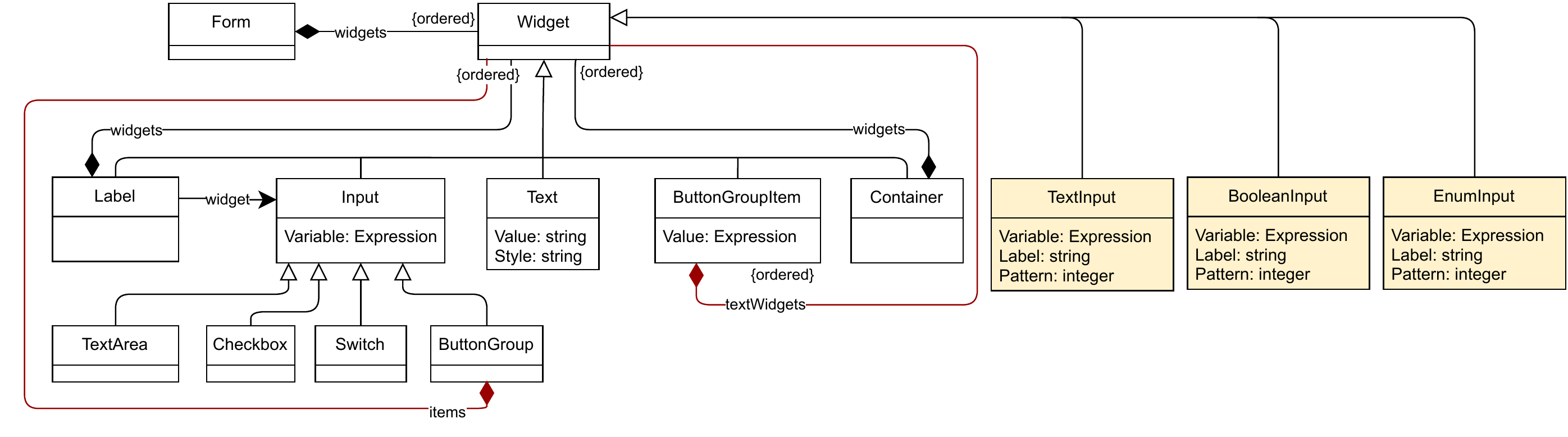}
    \caption{Virtual Widgets Metamodel}
    \label{fig:virtual_metamodel}
\end{figure*}

In \autoref{fig:virtual_metamodel} we present the metamodel for virtual widgets,
which is an extension of the native metamodel in \autoref{fig:native_metamodel}.
Whenever necessary and to avoid confusion, we refer to the concrete metamodel
and its widgets as \textit{native metamodel} and \textit{native widgets}.
In our example, the virtual metamodel includes three new classes\footnote{The new classes are depicted with a yellow background in the metamodel diagram.}: \widget{TextInput}, \widget{BooleanInput}, and \widget{EnumInput}. Each of these widgets represent a labelled input, and thus contain the \kw{Variable} and \kw{Label} attributes. Each element may correspond to one of several different concrete patterns, which are identified by the \kw{Pattern} property.
The virtual metamodel contains slightly modified versions of the classes in the native metamodel where aggregation relations are relaxed to refer to the \widget{Widget} base class instead of one of its subclasses\footnote{The modified relations are depicted in red.}.
For instance, compare the \widget{ButtonGroup} virtual widget
(\autoref{fig:virtual_metamodel}) with its native counterpart
(\autoref{fig:native_metamodel}). The next section explains in detail why this
change allows for the gradual transition between native and virtual models.

Taking advantage of the new virtual widgets \widget{TextInput} and \widget{BooleanInput}, we obtain a simplified virtual model (\autoref{lst:virtual_model}). A side-by-side comparison of the native and virtual widget trees as they would be displayed in an IDE  (\autoref{fig:trees}) further highlights the reduction in complexity.


\ \ \begin{minipage}{0.95\linewidth}
\begin{lstlisting}[basicstyle=\footnotesize\ttfamily, language=XML, caption={Form virtual model}, label={lst:virtual_model}]
<Form Id="f">
	<Container Id="c1">
		<TextInput Id="v1" Label="Request description" Pattern="1" Variable="Request.Description"/>
	</Container>
	<Container Id="c2">
		<BooleanInput Id="v2" Label="Approved by Manager" Pattern="3" Variable="Request.Approved"/>
	</Container>
</Form>
\end{lstlisting}
\end{minipage}

A virtual widget is defined by a list of properties and one or more patterns (see \autoref{fig:boolean_inputs_def}).
These patterns are used for representing virtual widgets in the virtual model, for viewing/parsing from native to virtual widgets, and for the synthesis of native widgets. 
Each pattern consists of a simplified native widget tree in which we only include the properties that must have a specific value for the pattern to be matched. Properties whose name is prefixed by \kw{Default} are used only during the synthesis phase and do not contribute to the pattern matching phase.
The pattern must also specify the binding of virtual widget properties to native widget properties.
The last line in \autoref{fig:boolean_inputs_def}, which provides a visual depiction of an instance of the pattern, was added for the convenience of the reader but is not part of the pattern definition.

In \autoref{fig:boolean_inputs_def} we present the definition of the
\widget{BooleanInput} virtual widget. This virtual widget declares two
properties: \kw{Variable} and \kw{Label}, and consists of three different
patterns. A property is made optional by providing a default value expression,
which is used exclusively during synthesis operations. In our example, the
default value for property \kw{Label} is obtained from the \kw{Value} property,
as specified by the \kw{GetLabelFor(Value)} expression\footnote{The
\kw{GetLabelFor} function calculates a suitable human-readable label from a
model expression.}.
Pattern \#1 consists of a \widget{Label} widget containing a \widget{Text} widget with identifier \kw{t}, followed by a \widget{Checkbox} widget with identifier \kw{i}.
The pattern specifies that the virtual property \kw{Variable} is bound to the checkbox's variable (i.e. \kw{Variable = i.Variable}), and the \kw{Label} property is bound to the text's value (i.e., \kw{Label = t.Value}).
Notice that we do not specify the \kw{Value} of any of the \kw{Text} widgets
because any value is acceptable\footnote{We do provided default values for two
of the \kw{Text} widgets of Pattern \#3 but these, as explained, do not affect
the pattern matching phase.}. However, for the two \kw{ButtonGroupItem} widgets
of Pattern \#3 we have explicitly specified their value to be \kw{true} and
\kw{false}, respectively, because we only want to recognize that pattern in that
particular case.

We check that patterns are well-formed with respect to the virtual widget definition (the pattern must provide
bindings for each of the virtual widget properties) and the native metamodel, in order to ensure that
the result of instantiating a virtual widget definition always results in correct native model.

This representation for virtual widgets is flexible enough to allow for easy and
safe model transformation. For instance, the native model may have been
initially defined with a set of widgets that match Pattern\#3, then parsed onto
the virtual model as a \widget{BooleanInput} virtual widget. The developer can
iteratively choose to use Pattern\#1 instead, which is finally
transformed back to the native model as a completely different set of native
widgets. This allows the user to construct an application in more abstract
terms, which allows for configuration steps that do not require complex weaving
of properties, values, and widgets.

\begin{table*}
    \caption{Boolean input patterns}
    \label{fig:boolean_inputs_def}
\centering
\begin{tabular}{l|l|l}
\hline
\multicolumn{3}{c}{\kw{BooleanInput(Expression {Variable}, string {Label} = {GetLabelFor(Value)})}}\\
\hline
    \textbf{Pattern\#1} & \textbf{Pattern\#2} & \textbf{Pattern\#3}\\
    \kw{Variable} = \kw{i.Variable} & \kw{Variable} = \kw{i.Variable} & \kw{Variable} = \kw{i.Variable}\\
    \kw{Label} = \kw{t.Value} & \kw{Label} = \kw{t.Value} & \kw{Label} = \kw{t.Value}\\
  \hline
    \begin{lstlisting}[basicstyle=\footnotesize\ttfamily,language=XML,numbers=none,frame=none,gobble=6,boxpos=t]
      <Label>
        <Text Id="t"/>
      </Label>
      <Checkbox Id="i"/>
    \end{lstlisting}
     & \begin{lstlisting}[basicstyle=\footnotesize\ttfamily,language=XML,numbers=none,frame=none,gobble=6,boxpos=t]
      <Label>
        <Text Id="t"/>
      </Label>
      <Switch Id="i"/>
    \end{lstlisting}
     & \begin{lstlisting}[basicstyle=\footnotesize\ttfamily,language=XML,numbers=none,frame=none,gobble=6,boxpos=t]
      <Label>
        <Text Id="t"/>
      </Label>
      <ButtonGroup Id="i">
        <ButtonGroupItem Value="true">
          <Text Default.Value="Yes"/>
        </ButtonGroupItem>
        <ButtonGroupItem Value="false">
          <Text Default.Value="No"/>
        </ButtonGroupItem>
      </ButtonGroup>
    \end{lstlisting}
  \\ \hline
    \includegraphics[width=.15 \linewidth]{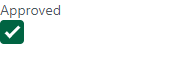}
     & \includegraphics[width=.15 \linewidth]{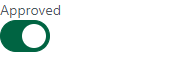}
     & \includegraphics[width=.15 \linewidth]{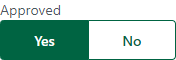}
  \\ \hline
\end{tabular}

\end{table*}

\section{The basic model parser}\label{sec:parsing}

\begin{figure*}
	\begin{small}
	\input{example/output/grammar}
	\end{small}
	\caption{Virtual Widget Grammar}
	\label{fig:grammar}
\end{figure*}

The virtualization process is accomplished using a parser-based approach. In
this work, we produce a parser generator that takes as input a grammar that
joins the virtual metamodel and a set of virtual widget definitions to produce a general
parser~\cite{tomita87} for concrete models. 
The input grammar for our running example is depicted in \autoref{fig:grammar}.
The grammar is mechanically created from the definition of the virtual
metamodel (\autoref{fig:virtual_metamodel}) and the definitions for virtual
widgets.
Notice that this context-free grammar contains
ambiguities between rules \widget{BooleanInput.3} and \widget{EnumInput.2}.
Given the grammar ambiguity, it is necessary to use a parsing algorithm that
explores all possible derivation trees of a word~\cite{Knuth65}. We adapted the
parser generator algorithm proposed by Tomita~\cite{tomita87} which provides an
efficient parsing algorithm for this kind of grammar.

In the grammar definition presented in \autoref{fig:grammar}, the non-terminal symbols
are represented in italics. The terminal symbols represent native model
instances and are denoted by XML tags, and have an optional subscript identifier and an optional superscript
condition.
Besides the structure of the main elements of the metamodel, the grammar
includes rules for each one of the patterns of each virtual widget (rules
\widget{TextInput.1}, \widget{BooleanInput.1}, \widget{BooleanInput.2},
\widget{BooleanInput.3}, and \widget{EnumInput.1}), as well as rules for
each of the extended native widgets.

The rules for virtual widgets are directly obtained from their definition. The definitions of \autoref{fig:boolean_inputs_def} lead to rules \widget{BooleanInput.1}, \widget{BooleanInput.2}, and \widget{BooleanInput.3}. 
For simplicity, we have omitted the definitions of the \kw{TextInput} and \kw{EnumInput} virtual widgets but included the corresponding rules in the grammar.
Patterns can capture the identifiers of model widgets they match by providing an \textit{Id} attribute, which corresponds to the terminal symbols subscripts in grammar rules, and bind it to variables in the semantic actions. 
Property value restrictions result in conditions that need to be met during
parsing. Conditions are represented as superscripts in the grammar. For
instance, in the \widget{BooleanInput.3} rule the first occurrence of the
\kw{<ButtonGroupItem>} terminal is annotated with \kw{Value = "true"}. When
parsing, if the condition is not satisfied by the current token then the rule is
immediately rejected.
If no other rule is active at that point, the parser will backtrack to the previous most recent decision point.
This approach is similar but less restrictive than the predicates of pred-LR(k) parsers \cite{parr1993use}, which requires exactly one predicate to true when there is more than one predicate in the currently active rules.

The grammar also includes rules for (extended) native widgets to enable the gradual extension of the virtual widget set. This ensures that every native model can be parsed into a virtual model, and the patterns that match the virtual widgets already defined are abstracted.
Looking at our virtual model example (\autoref{lst:virtual_model}), it can be seen that we have a mix of extended native widgets (\widget{Form} and \widget{Container}) and virtual widgets (\widget{TextInput.1} and \widget{BooleanInput.3}) in the same model.

The patterns in the definition of virtual widgets are allowed to overlap. In conjunction with the inclusion of rules for the native widgets, this leads to grammars with a high degree of ambiguity. As an example, consider the widget sequence of lines 9 to 19 of \autoref{lst:native_model}. This sequence can be parsed in several different ways, namely using either rule \widget{BooleanInput.3}, \widget{EnumInput.1}, or \kw{<Label>} followed by \kw{<ButtonGroup>}.
As mentioned before, we follow an approach inspired by Tomita's GLR parsers~\cite{tomita87}. While Tomita explores all parsing trees in parallel, we keep an active parse tree at each time but remember past decisions in shift-reduce and reduce-reduce conflict situations, and thus allow backtracking. Notice that our goal is to find a parse tree as quickly as possible instead of finding all possible parse trees.
According to our benchmark (detailed in \autoref{sec:evaluation}), backtracking happens at a rate of 2.6\% per processed token which in pratice leads to the parser converging quickly.
When encountering a shift-reduce conflict we explore the shift route first and thus give priority to longer sequences. For reduce-reduce conflicts, we explore the higher priority rule first, according to an explicit order fixed when defining the virtual widgets.
Additionally, rules for virtual widgets have higher priority than rules for native widgets.

The semantic action for each virtual widget rule (which we are not representing in \autoref{fig:grammar}) amounts to creating an instance of the corresponding virtual widget. The properties for the instance are extracted from the native widgets according to the pattern property bindings.
E.g., rule \widget{Boolean.3} is used to reduce the widget sequence of lines 9 to 19 of \autoref{lst:native_model}.
Given the property bindings specified in Pattern \#3 (\autoref{fig:boolean_inputs_def})
we obtain the following mapping between terminals and model widgets: $\{ \kw{t} \mapsto \kw{t1}, \kw{i} \mapsto \kw{bg} \}$. 
Moreover, this association results in the following values for the virtual widget properties (line 6 of \autoref{lst:virtual_model}):
\begin{itemize}
\item \kw{Variable} = \kw{Request.Description}
\item \kw{Label} = \kw{"Request description"}
\end{itemize}

The value for \kw{Variable} is calculated as follows:
\begin{enumerate*}
\item Pattern\#3 specifies the binding \kw{Variable} = \kw{i.Variable},
\item pattern widget \kw{i} is matched with native widget \kw{bg},
\item replacing \kw{i} by \kw{bg} in the binding we \kw{Variable} = \kw{bg.Variable}
\item the value of \kw{bg.Variable} is \kw{Request.Description}.
\end{enumerate*}

Given the special circumstances of processing a concrete low-code model and having special
conflict resolution policies and backtracking, we have developed our own GLR parser
generator. The characteristics of our specific use case that made it
difficult to directly use any of the widely available parser generators are the following:
\begin{itemize}
\item Grammar ambiguity: most parsers require the grammar to be non ambiguous. As explained above, our grammars are always ambiguous, by design.
\item Ambiguity resolution and backtracking: shift-reduce and reduce-reduce conflicts are not solved in a single way. Instead, they represent alternatives that are explored in sequence and from which we may need to backtrack.
\item Adapting to new virtual widget definitions: our grammar and the corresponding parser are generated on the fly (in memory). When a model is loaded in the IDE the most recent definitions are fetched.
\item Conditions in terminal symbols: property value restrictions in the patterns are represented as conditions in the associated terminal symbols. Hard restrictions on the parsing time demand these conditions to be evaluated as soon as possible, instead of waiting until the semantic action (potentially) rejects the rule.
\end{itemize}

The inclusion of conditions in terminal symbols is crucial to the conciseness of
our grammars and the efficiency of the parser. Conditions could be alternatively
represented by cloning the relevant terminal symbols. E.g., given the existing conditions
for  terminal \kw{ButtonGroupItem}, we can split it into three
different terminals \kw{<ButtonGroupItem\_true>}, \kw{ <ButtonGroupItem\_false>}, 
and \kw{<ButtonGroupItem\_other>} (plus the corresponding close tags),
representing the three relevant cases for the value of the \kw{Value} property:
true, false, and any other expression.
Rule \widget{BooleanInput.3} can be readily converted to use the new terminal
symbols, but rule \widget{ButtonGroupItem} needs to be split into three different
rules, each using one of the new terminals. Cloning terminal symbols would lead
to a combinatorial explosion of the grammar, and an exponential increase in the
number of states in the parser.


Notice that by construction the conditions on terminal symbols are always of the
form \textit{property = value}. Our parser generator takes advantage of this
fact to efficiently generate the parser tables. An entry in the parser table
specifies the actions to take (shift / reduce / none) for each terminal /
non-terminal symbol based on the current token. In our representation, each
action may include a decision table like the ones depicted in
\autoref{fig:parser_table} showing a fragment of the parser table for our
example grammar.
When in state 32 of \autoref{fig:parser_table} we have processed the sequence

\kw{<Label> <Text/> </Label> <ButtonGroup>}

\noindent
At this point, and according to the parse table, the only
expected tokens are \kw{</ButtonGroup>} and
\kw{<ButtonGroupItem>}\footnote{Any other token results in a parsing error and the parser
backtracking to the most recent decision point.}. The first token results in a
reduction action using rule \widget{Enum.Input.1} and the creation of an \widget{EnumInput} virtual widget. The second token is more
interesting. Recall that the \widget{ButtonInput.3} rule has attached the condition
\kw{Value = "true"} for the first \widget{ButtonGroupItem}.
The parser table in \autoref{fig:parser_table} captures this information by
specifying that a shift will happen and that the target state is determined by
the value of the \kw{Value} property: if the value is \textit{true} it shifts to
state 37, whereas any other value shifts to state 36.


\begin{table}
\caption{A fragment of the parser table}
\label{fig:parser_table}     
     \small
\begin{tabular}{rll}
\toprule
 & \multicolumn{2}{c}{Token}\\
\toprule
     State & $<\!\!/$ButtonGroup$>$ & $<$ButtonGroupItem$>$ \\
\midrule
     32 & Reduce(EnumInput.1) &  Shift(Value:\{ true $\mapsto$ 37, \\
     									& & \qquad \qquad \quad \ \ other $\mapsto$ 36 \}) \\
     48 & Reduce(EnumInput.1) &  Shift(Value:\{ false $\mapsto$ 50, \\
     									& & \qquad \qquad \quad \ \ other $\mapsto$ 36 \}) \\
\bottomrule
\end{tabular}

\end{table}


\section{Native model synthesis}\label{sec:synthesis}

The virtual widget definitions are twofold, as depicted in
\autoref{fig:boolean_inputs_def}: they define the patterns used to produce the
parser that virtualizes the native model and instantiates virtual widgets, and
define how each virtual widget is mapped back to the native model, in this case,
assigning several alternative mappings to the same virtual widget.
%
%
The virtual model is transformed back into the native model (i.e. native model
must be \textit{synthesized}) in the following circumstances:
\begin{enumerate*}
\item instantiation of virtual widgets,
\item switching the pattern of a virtual widget,
\item and changing the value of a virtual widget property
\end{enumerate*}.

Definitions of virtual widgets contain a set of property definitions, which are
bound to the properties of native widgets using annotations in the native
widget patterns and equations in each pattern to relate them. 
In the case of the virtual widget \texttt{BooleanInput} defined in
\autoref{fig:boolean_inputs_def}, the parameters are \kw{Variable} of type
{Expression} and \kw{Label} of type string. 
Equations in this definition can be read in both ways -- from the virtual widget
to the native widget and vice-versa.
During the virtualization phase, the equation $\kw{Variable}=i.\kw{Variable}$
in Pattern\#1 means that the value of the \kw{Variable} property of the virtual
widget is set with the value of the \kw{Variable} property of the native widget
\kw{Checkbox} identified by $i$.
During synthesis (which occurs when new virtual widget instances are created
or the developer changes the pattern of an existing virtual widget), the same equation
specifies that the value of the \kw{Variable}
property of the (newly created) native widget \kw{Checkbox} identified by $i$ is set with the value
of the \kw{Variable} property of the virtual widget.

The equations provide the value for some of the properties of the synthesized
native widgets. The remaining properties are set if they have been specified in the
pattern, either as matching values or default values. For instance, when instantiating Pattern\#3
of \widget{BooleanInput}, the \kw{Value} for the first \kw{ButtonGroupItem} widget
is set to \kw{true}, and the \kw{Value} of its \kw{Text} widget is set to \kw{Yes}.

Additionally, all native widget properties are preserved when changing patterns if there
is a corresponding widget in the target pattern.
For instance, consider the virtual model \autoref{lst:virtual_model} and changing
the \widget{BooleanInput} instance with id \kw{v2} from Pattern\#3 to Pattern\#1.
The widget with id $t$ in Pattern\#3 corresponds to the native widget with
id $t3$ (\autoref{lst:native_model}). Pattern\#1 also contains a widget with id $t$.
Consequently, the synthesized \kw{Text} widget will have a non-empty value for
property \kw{Style} even though the pattern provides no value for the property.
\autoref{lst:transformed_model} displays the resulting native model.

\ \ \begin{minipage}{0.95\linewidth}
\begin{lstlisting}[basicstyle=\small\ttfamily, language=XML,firstline=9,lastline=12,caption={Result of changing \widget{BooleanInput} from Pattern\#3 to Pattern\#1 (snippet)}, label={lst:transformed_model}]
		<Label Id="l2" Widget="bg">
			<Text Id="t2" Style="bold" Value="Approved by Manager"/>
		</Label>
		<Checkbox Id="bg" Variable="Request.Approved"/>
\end{lstlisting}
\end{minipage}

Of course, this bidirectional reading capability limits the kind of equations
that can be used in the definition. At this point, we limit equations to directly
reading/setting properties of virtual widgets and native widgets.


\section{Evaluation}\label{sec:evaluation}



We designed a performance benchmark with a set of large examples to test the
efficiency and effectiveness of our method.
In this batch, we used the real OutSystems metamodel (consisting of 30 native widget classes) augmented with $31$ virtual widget definitions and $106$ patterns in
total, which corresponds to the totality of the gradually defined database of
patterns at the time of testing. We used this set of virtual widget definitions
to further assess the correctness of the parser generator process and also to
sample the correctness of the synthesis process that consisted of switching
between patterns.
We ran a warmup process and repeated the experiments $9$ times for each module,
to drop the $2$ highest and the $2$ lowest values, thus reducing noise, and obtaining the parsing time in isolation. 
As can be seen in \autoref{fig:parser_statistics}, the generated parser is highly ambiguous.

\begin{table}
\caption{Generated parser statistics for the OutSystems metamodel augmented with 31 virtual widgets}
\label{fig:parser_statistics}     
     \small
     \centering
     
\begin{tabular}{lr}
	\toprule
	Grammar rule count & 188 \\
	Parser state count & 926 \\
	Shift-reduce conflicts count & 216 \\
	Reduce-reduce conflicts count & 982 \\
	\bottomrule
\end{tabular}
\end{table}

\begin{figure*}[t]
\centering
 \includegraphics[width=0.58\linewidth]{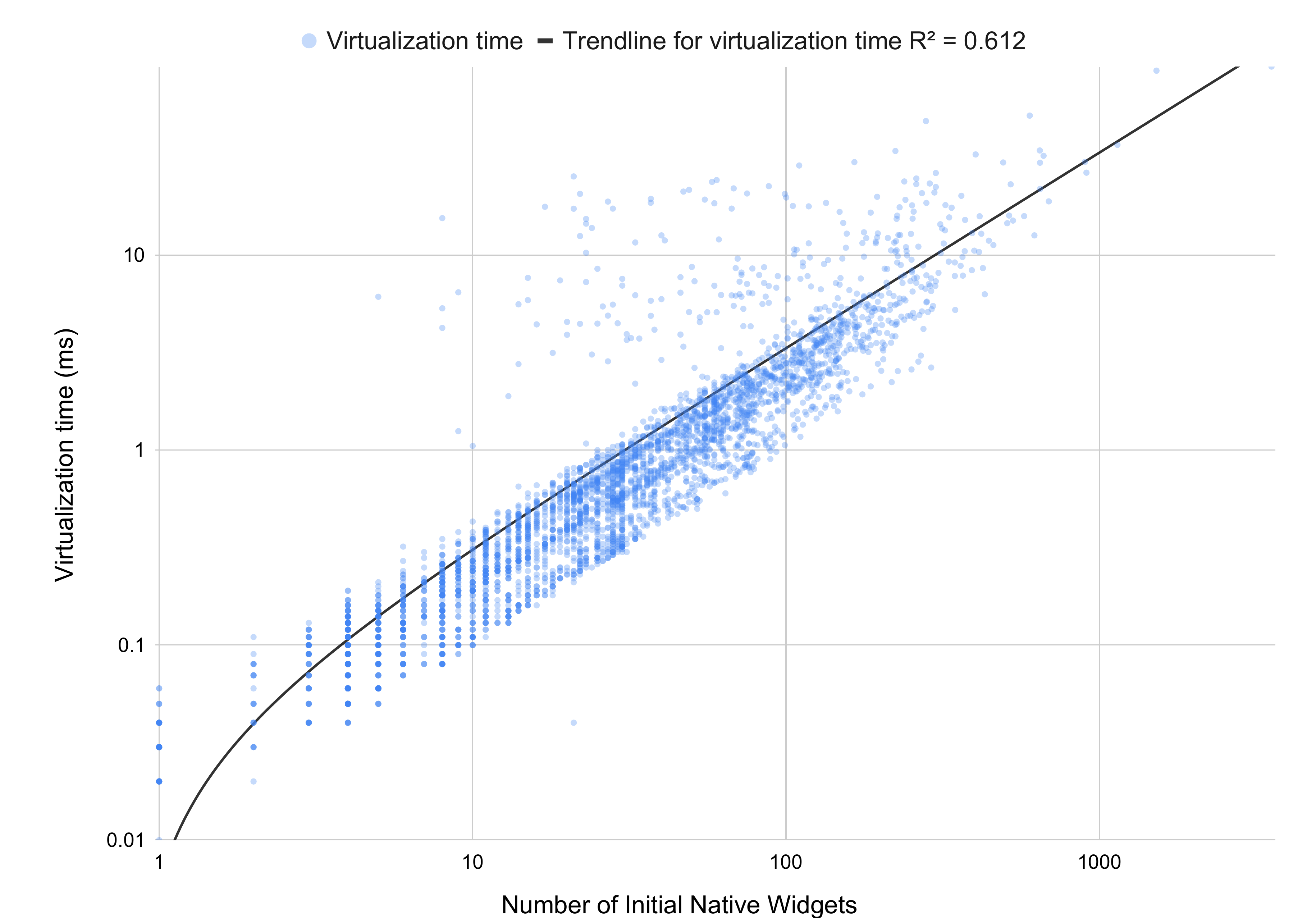}
\caption{Parsing time per number of native widgets (log-log)}
\label{fig:widgetPerSecond}
\end{figure*}

To assess the efficiency and effectiveness of our approach we designed a simple
experiment to use the virtual widget definitions in a large code base. Our
sample consists of $239$ OutSystems modules containing $3~718$ screens, summing
up a total of $198~072$ native widgets.
\autoref{fig:widgetPerSecond} shows the virtualization (parsing) time per number
of native widgets in every module using a double logarithmic scale, to allow
depicting small and large applications simultaneously.
We found a clear linear trend ($t = 0.0336*n-0.0277$) in the parsing time in
milliseconds ($t$) with a $33.6\mu s$ increase in parsing time for every new
widget parsed ($n$).
The coefficient of determination of the linear regression ($R^2=0.611$), with
$\mathit{p{-}value} < 0.001$, shows that there is a good fit, but there are
still several cases where the time is not linearly related to the number of
native widgets. The chart of \autoref{fig:widgetPerSecond} shows that about $14\%$ of cases are slower than the
trendline, which is visible by the existence of outliers visible in the chart. 
In our raw data, we observe that the longest parsing time for a module is about
$92ms$, which is still within the range of acceptable times for an interactive
system, where $100ms$ is the reference number for the time limit for a user to
feel that they are manipulating objects live in a user
interface~\cite{10.1145/108844.108874}. In most cases, the time is quite short,
and the reasons for the outliers are related to extreme cases of ambiguity that
triggers excessive backtracking in the parsing process. Most of the more direct
ambiguous cases are immediately dealt by the GLR parsing algorithm that is set
to select the longest sequence of native widgets.


\begin{figure*}[t]
  \centering
  \includegraphics[width=0.58\linewidth]{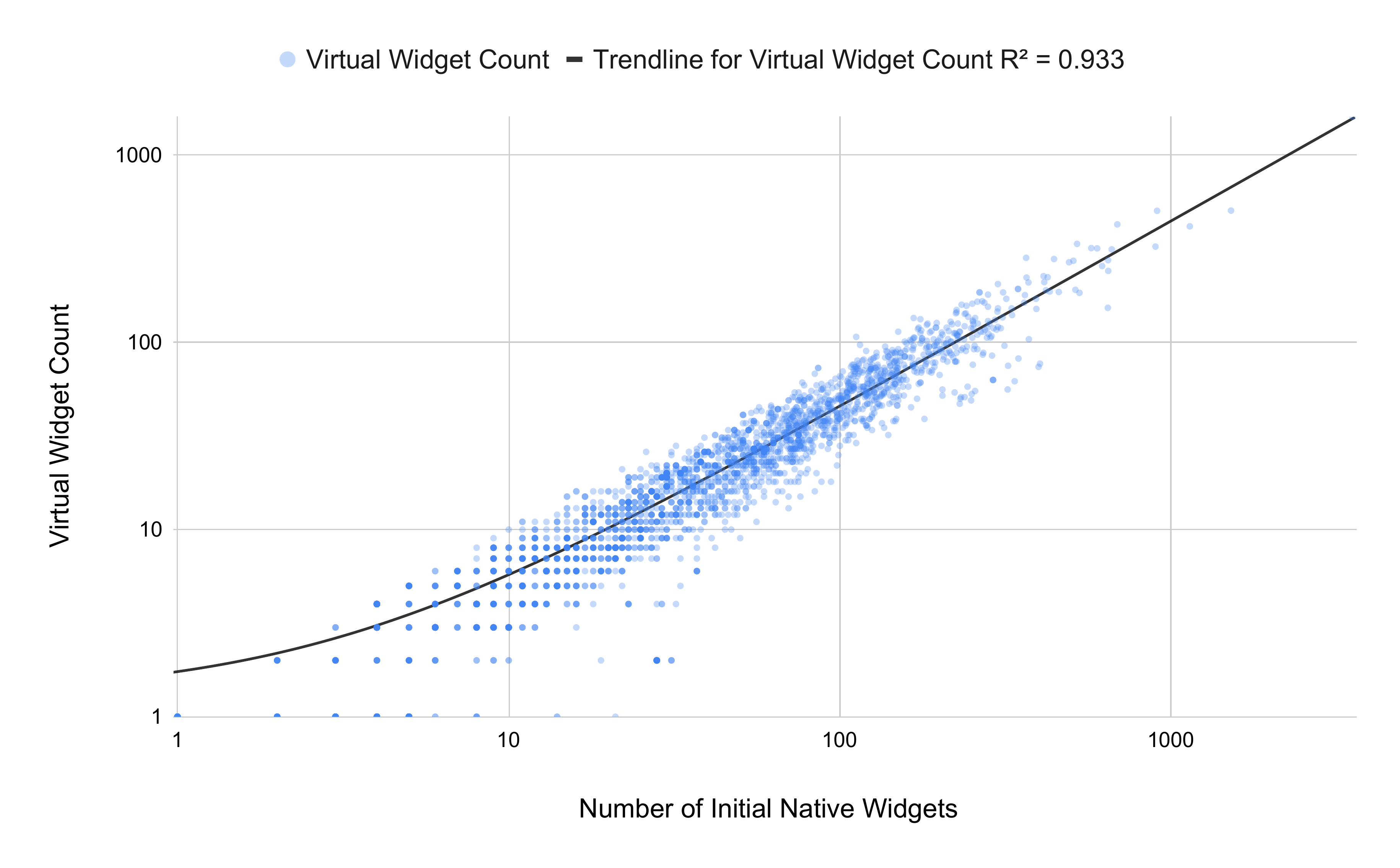}
  \caption{Number of virtual widgets parsed per number of initial native widgets (log-log)}
  \label{fig:vwidgetPerNWidget}
\end{figure*}

As far as effectiveness is concerned, we observe, in
\autoref{fig:vwidgetPerNWidget}, that the average number of widgets that is
abstracted ($v$) is even steadier with a linear trend ($v = 0.444*n + 1.3$) with a
high coefficient of determination ($R^2 = 0.977$) and $\mathit{p{-}value} < 0.001$. The
ratio is mostly constant and close to $1{:}2$, which means that on average the
number of widgets is reduced to half in the abstract model with only 31 virtual widgets definitions. It is expected that
this ratio increases with the gradual addition of new virtual widgets definitions.

As depicted in \autoref{fig:parser_statistics}, the parser that we generated for our benchmark is highly ambiguous. The large number of reduce-reduce conflicts is due to the overlap of virtual widget and native widget rules. However, the high degree of ambiguity has a very low practical impact on the parser performance.
The statistics of \autoref{fig:parsing_statistics} back our claim.
The parser processes a total of 307~884 tokens when running our full benchmark. In an ideal scenario where backtracking never occurs, an equal amount of shift operations is expected, i.e. each token is processed exactly once. As can be seen, the total number of shift operations is higher (318~515), but not by a large margin - on average the parser runs 1.035 shift operations per input token.
Conflicts are encountered at 7.95\% of the tokens. Notice that a conflict corresponds to the availability of multiple parser actions for the same lookahead token, which our parser handles by prioritizing one of the rules and keeping track of the remaining ones in case it needs to backtrack.
Bactracking occurs at a lower rate of 2.57\%. A lower value is expected because not all conflicts lead to backtracking.
The impact of backtracking is low. A total of 10~631 of tokens had to be backtracked, i.e. reprocessed. On average, backtracking results in going back 1.341 tokens.

In summary, these results show that our gradual approach is suitable for a live
bidirectional transformation of models that allows simultaneous editing of
application models at different levels of abstraction without overriding or
losing any detail in the model visualization or synthesis. 

\begin{table}
\caption{Benchmark parsing statistics}
\label{fig:parsing_statistics}   
     \small
\begin{tabular}{lr}
\toprule
	Total input token count & 307 884 \\
	Shift operations count & 318 515 \\
	Total conflict count & 24 477 \\
	Backtracking operations count & 7 927 \\
	Total backtracked tokens & 10 631 \\
\midrule
	Average shift operations per input token & 1.035 \\
	Conflict rate per input token & 7.95\% \\
	Backtracking rate per input token & 2.57\% \\
	Average backtracked tokens per backtracking operation & 1.341\\
\bottomrule
\end{tabular}  
\end{table}


\section{Related Work}\label{sec:relatedwork}
 
Bidirectional transformations have been applied in a variety of
contexts~\cite{DBLP:conf/icmt/CzarneckiFHLST09}. For instance, in the database
community there is an extensive body of work regarding the view-update
mechanism~\cite{10.1145/319628.319634, DBLP:conf/vldb/Masunaga84,
10.1145/49346.50068, DBLP:conf/pods/BohannonPV06}. Hence, there are multiple
scenarios where a focus on the efficiency of these transformations is crucial.

In the context of Model-Driven Engineering (MDE) bidirectional transformations
play a critical role~\cite{kahani2019,guerra2010} where it is the cornerstone to
define flexible Model-to-Model transformations (M2M).
Guerra et al.~\cite{guerra2010} argue that M2M transformations are usually
expressed through specialised M2M transformation languages and are mostly
focused on the implementation phase. Their work~\cite{guerra2010} addresses this
issue by proposing a family of languages to cover the full transformation
development life cycle. Our approach can be seen as an instance of the
methodology proposed in~\cite{guerra2010} as we cover the whole lifecycle phases
of the transformation development in our tool, from the requirements to the
design and validation. By providing a back office and a metamodel to define
virtual widgets, we are enabling pattern designers to extend the transformation,
and by providing validations in this editor we are ensuring that the produced
transformation is correct every step of the way.

Comparing \chameleon with M2M approaches as EMF~\cite{SteinbergBudinskyEtAl09}
or MPS~\cite{MPS,10.1145/2500828.2500846} the disadvantage is that those
approaches define transformations as external to the language itself (\cf
ATL~\cite{DBLP:journals/scp/JouaultABK08,Atlas}). \chameleon uses the native metamodel, extended with name
binders and equations to express the patterns for virtualization and synthesis.

Triple Graph Grammars (TGGs)~\cite{TGGs,OverviewTGGs} are a rule-based approach
proposed to define transformations between graph-based languages while
maintaining the bidirectional relationships between input and output elements.
TGGs are particularly useful to synchronise different models and determine the
compatibility of source and target models, which in \chameleon is (mostly)
ensured by construction as the metamodel for virtual widgets subsumes the
metamodel for native virtual widgets. But a more critical issue with TGGs is
efficiency, as parsing algorithms for TGGs are complex and naive solutions may
lead to an exponential analysis~\cite{OverviewTGGs}. Even by constraining TGG
expressivity, at best, dedicated parsers have polynomial complexity. In
contrast, \chameleon has a linear parsing algorithm, which is possible since it
deals with abstract syntax trees. In the low-code setting of \chameleon there
are graph-based structures similar to flow charts (action flows) that can
nevertheless be expressed as structured (tree-like code) and thus processed
efficiently.

Viatra~\cite{viatra,viatra3} uses formal mathematical techniques of graph
transformation and abstract state machines. This framework can perform complex
transformations by using an incremental pattern matching approach that enables
the determination and updating of matches after modifications are made
incrementally. Nevertheless, the execution time of the transformations can be
significant for large models, hindering their use in interactive systems.

QVT~\cite{QVT,DBLP:conf/models/Stevens07,DBLP:journals/sosym/Stevens10} is a
family of MDE languages that provide a standard for (mainly) defining model
transformations between different types of models. QVT relations~\cite{QVT}
supports complex object pattern matching and object template creation with
correspondences between domains are specified as relations. In particular, QVT
relational uses a strategy where it applies simultaneously all matches of
translation rules to a given input model and then merges the resulting output
model elements. However, this approach tends to be prone to errors and demands a
deep understanding of how the rules match and interact with each
other~\cite{tuprints2894}. As a result, available QVT relational tools can yield
significantly different outcomes when processing the same
input~\cite{DBLP:journals/sosym/GreenyerK10}.

The works~\cite{DBLP:conf/fase/HermannEEO12, DBLP:journals/sosym/XiongSHT13}
address concurrent model synchronization, where 
transformations
may be applied simultaneously to source and target models. Currently \chameleon
does not support this feature, but we are exploring how to include concurrent
model synchronization in our solution.

\section{Future Work}\label{sec:futurework}

While \chameleon represents a promising approach to model-driven engineering, it
is important to note that it is not a fully general solution. Specifically,
\chameleon does not cover models that are intrinsically graph structures as these
scenarios do not typically arise in the OutSystems context. 

One of the constraints we imposed in this work is that the model only uses  
simple parser conditions on enumeration values. Although our implementation is
capable of handling general conditions, we found that this restriction can lead
to a combinatorial explosion of states in the parser, which hinders the
interactivity of the system. However, in practice, this restriction has
shown it is not a limitation in the definition of virtual widgets.

In the present proposal, one open issue that may influence the results is the
amount of backtracking that occurs during the parsing process. In the future, we
plan to deeply study the cases where backtracking is necessary and find ways to
avoid or optimize it.

We also identify possible extensions of this work to integrate it further. One
of the most important extensions is to allow the definition of virtual widgets
using prior developed template language \ostrich~\cite{models2021,models2022,SoSym2022}.
The visualization of the model using templates allows for greater reuse of code
and links this system with a database of curated patterns that already exist in
the platform. Also, the binding mechanism would benefit from using template
parameters instead of the direct binding of names to the model.

\section{Conclusions}\label{sec:conclusions}

In this work, we studied the application of standard parsing techniques to
implement model abstraction and the definition of linked native patterns to
define a virtual widget. Virtual widgets, defined in a metamodel as an extension
of the native metamodel, lead to a simplified visualization of the model in
widget trees without discarding any part of the model. 

By allowing the incremental definition of the correspondence between virtual
widgets with native widget patterns and name binders with equations, we support
a gradual approach to the virtualization of a low-code metamodel thus raising
even further the abstraction level of the native language. 

The virtual metamodel we introduce is dynamic, produced from a set of virtual
widget definitions, and then instantiated in an efficient model parser. 
We evaluated the resulting system according to efficiency and effectiveness
parameters using a significantly large benchmark of real (low) code. All in all,
the approach is shown suitable for collaborative and interactive editing of the
low-code at different abstraction levels simultaneously. 


\bibliographystyle{IEEEtran}
\bibliography{IEEEabrv,main}

\end{document}